%% file: main.tex
\documentclass[10pt, conference, letterpaper]{IEEEtran}

\usepackage[T1]{fontenc}

\usepackage{cite}
\usepackage{textcomp}           
\usepackage{color}
\usepackage{subfigure}
\usepackage{booktabs}
\usepackage{hyperref} 

\usepackage{enumitem} 

\usepackage{balance}

\usepackage{xspace}
\usepackage{microtype}
\usepackage[british]{babel}
\usepackage{graphicx}
\graphicspath{{graphics/}}

\definecolor{orange}{rgb}{.9,.4,.1}
\newcommand{\todo}[1]{\textcolor{red}{TODO: #1}}
\newcommand{\oleg}[1]{}

\newcommand{\comment}[1]{}

\begin{document}
\title{A Case for Time Slotted Channel Hopping\\ for ICN in the IoT}


\author{\IEEEauthorblockN{Oliver Hahm\IEEEauthorrefmark{1},
C\'edric Adjih\IEEEauthorrefmark{1},
Emmanuel Baccelli\IEEEauthorrefmark{1},
Thomas C. Schmidt\IEEEauthorrefmark{2},
Matthias W\"ahlisch\IEEEauthorrefmark{3}}
\IEEEauthorblockA{\IEEEauthorrefmark{1}INRIA \hspace{.2cm}
\IEEEauthorrefmark{2}HAW Hamburg \hspace{.2cm}
\IEEEauthorrefmark{3}Freie Universit\"at Berlin}
\IEEEauthorrefmark{0}{\{oliver.hahm,emmanuel.baccelli,cedric.adjih\}@inria.fr,
t.schmidt@haw-hamburg.de, m.waehlisch@fu-berlin.de}
}

\maketitle


\begin{abstract}
Recent proposals to simplify the operation of the IoT include the use of Information Centric Networking (ICN) par\-adigms.
While this is promising, several challenges remain.
In this paper, our core contributions (a) leverage ICN communication patterns to dynamically optimize the use of TSCH (Time Slotted Channel Hopping), a wireless link layer technology increasingly popular in the IoT, and (b) make IoT-style routing adaptive to names, resources, and traffic patterns throughout the network---both without cross-layering.  
Through a series of experiments on the FIT IoT-LAB interconnecting typical IoT hardware, we find that our approach is fully robust against wireless interference, and almost halves the energy consumed for transmission when compared to CSMA. Most importantly, our adaptive scheduling prevents the time-slotted MAC layer from sacrificing throughput and delay.   


%
%
\end{abstract}


\input{text/intro}
\input{text/problem_statement}

\input{text/related}

\input{text/icn-over-tsch}
\input{text/scheduling}
\input{text/evaluation}

\begin{small}
\balance
\bibliographystyle{IEEEtran}
\bibliography{2015-icn-tsch}
\end{small}
\end{document}

%% file: text/intro.tex
\section{Introduction}\label{sec:intro}

The current Internet is based on IP as convergence layer, and focuses primarily on the interconnection between machines--- the byproduct of which being that these machines can then store, send and receive digital content. 
ICN proposes a shift towards a simplified convergence layer focusing directly on digital content access and distributed storing.
ICN is considered both (i) as a \emph{clean-slate approach}, running on top of the MAC layer, and (ii) as an \emph{overlay approach}, running on top of the IP stack.

An interesting domain in which ICN is being studied as a clean-slate approach is the Internet of Things (IoT), e.g. in \cite{baccelli+:2014}. 
The IoT has already started being deployed, and will consist in large part in the interconnection of tens of billions of resource-constrained communicating devices \cite{rayes2012internet}, e.g. smart sensors and actuators of various kinds, a.k.a \emph{Things}. 
This deployment is expected to generate massive amounts of data that will both (i) allow the optimization of existing processes, e.g. large-scale complex industrial processes, and (ii) support entirely new mechanisms and businesses based on the simultaneous availability of this data and ever increasing environment automation. 

In this paper, we will study aspects of clean-slate ICN approaches, applied in IoT scenarios such as industrial Internet \cite{RFC-5673}. 
In this context, ICN approaches are studied and experimented with in the hope that they can solve several hard problems at once, including end-to-end security, and significantly increased energy efficiency, while fitting much tighter on-device memory constraints. 
However, lessons learnt so far in the IoT point towards conflicting requirements concerning MAC layers. 
On one hand, wireless communications are mandatory to provide the necessary cost-effectiveness and flexibility prohibited by the deployment and maintenance of too many wires. 
On the other hand, wireless communications are typically plagued with drastic reliability issues, compared to wired communications. 
In this paper, we thus focus on the interplay between wireless MAC layers and ICN mechanisms. 
We show how ICN characteristics can benefit from and optimize the use of novel link layers based on combinations of time-division multiple access and frequency hopping.

The remainder of this paper is organized as follows. 
\S~\ref{sec:problem} reviews related work in the domain of ICN, and states the problem we focus on: matching ICN paradigms and IoT link layer characteristics.
\S~\ref{sec:related} recalls the main IoT techniques relevant for this paper, including background on TSCH.
\S~\ref{sec:icn-tsch} provides an overview of the architecture we propose for optimized operation of ICN over a TSCH-based wireless link layer.
\S~\ref{sec:scheduling} focuses on designing approaches for efficient scheduling of ICN Interest/Chunk traffic.
\S~\ref{sec:evaluation} evaluates these approaches on typical IoT hardware in a testbed, with an implementation based on NDN, RIOT \cite{RIOT}, and 802.15.4e (OpenWSN \cite{OpenWSN}).

%% file: text/problem_statement.tex
\begin{figure*}[t!]%
\centering%
\includegraphics[width=\linewidth,natwidth=100,natheight=100]{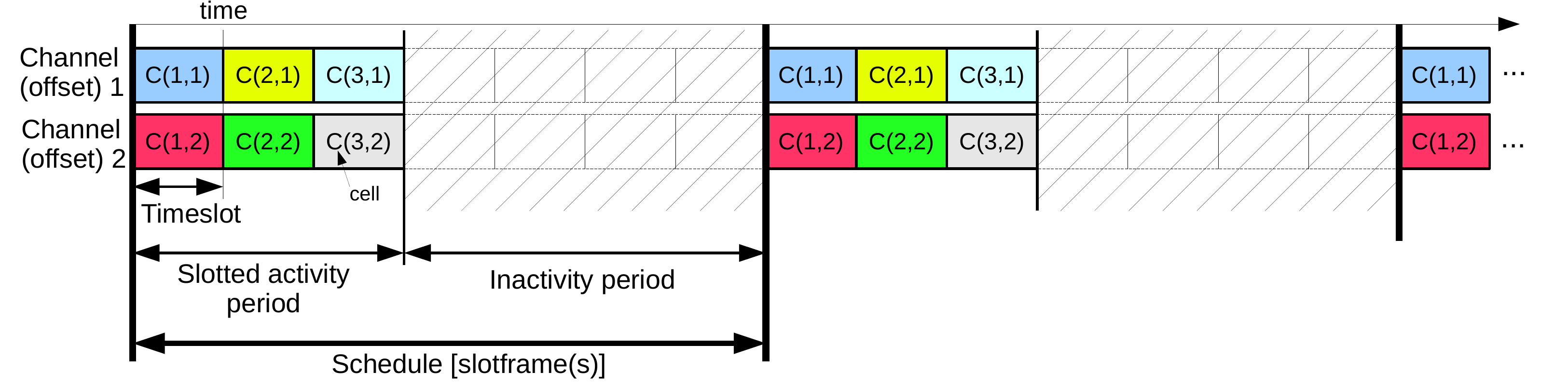}
\caption{
Time-Division Multiple Access (TDMA) represented horizontally, Time Slotted Channel Hopping (TSCH) represented vertically.
}%
     \vspace{-1.0em}
\label{fig:tdma-base}%
\end{figure*}

\section{ICN and IoT: The State of Affairs}\label{sec:problem}

The IoT will include a large number of devices with constrained resources~\cite{RFC-7228}, which communicate via wireless channels. 
Deploying ICN in this context may not only facilitate some applications, but also simplify the protocol complexity and increase network efficiency \cite{draft-irtf-icnrg-challenges,baccelli+:2014}. 
However, wireless transmission between IoT devices is typically built on contention-based MAC protocols (CSMA is the archetype), which are unreliable, prone to collisions and high packet loss. When combined with ICN approaches, such as NDN on which we focus in this paper, this unreliability leads to a number or problems, described~below.

\subsection{Problem Statement}

Significant unreliability with wireless MAC protocols impedes the generation of coherent pending Interest paths, and amplifies the problem of state de-correlation \cite{wsv-bdpts-13} of NDN stateful reverse path forwarding (RPF).

Retransmissions and overload---as well as de-localized content in an unknown topology---can add high latencies to the information centric request-response pattern and lead to unpredictably high RTT fluctuations\cite{wsv-bdpts-13}. 
For NDN in fluctuating wireless environments, a node cannot reliably estimate when it will receive a content chunk in response to a pending Interest, or when it should retransmit this Interest.
A forwarded Interest may have been lost or delayed somewhere in the network, or the Interest was just not satisfiable anywhere in the network.
Consequently, it is hard to set a reasonable timeout for retransmitting the Interest, without any knowledge about transmission delays. 

A further concern lies in energy consumption due to (avoidable) activity over radio. 
Aside from error recovery, receiver capacities are quickly drained by excessive broadcasts that occur from frequent reconfigurations, or unsophisticated routing practice. 
A particular problem in the wireless IoT domain thus lies in lightweight re-configurations and seamless route acquisitions. 
The latter poses a specific challenge, since the space of named routable entities is particularly large in the ICN world \cite{draft-irtf-icnrg-challenges}.

The typically unreliable and fluctuating nature of wireless communication in the IoT thus has a strong impact on the functionality of an ICN layer. 
This motivates the search for an alternative link layer technologies, which allow more appropriate cooperative use of the radio. 
A promising candidate is Time Slotted Channel Hopping (TSCH)~\cite{802-15-4-2011,802-15-4e-2012}, which replaces CSMA with a reservation-based MAC protocol, combining TDMA with frequency hopping. 

TSCH can drastically increase the reliability of packet transmission~\cite{watteyne+:2015} thereby guaranteeing a fixed throughput and maximum latency even at high traffic load---if a proper schedule exists.
However, an a priori derivation of a schedule requires thorough understanding of future traffic flows in the network which is infeasible for most application domains. 
Furthermore, traffic patterns and profiles may vary over time, leading to largely fluctuating demands that contradict the approach of a static schedule. 
In general, TSCH allows for a dynamic slot scheduling, but schedule negotiations are expensive. 
An approach for improving wireless ICN by TSCH thus poses the challenging problem of deriving and maintaining an adaptive scheduling of communication slots at an affordable cost.


\subsection{Related Work}

ICN has been identified as potential key enabler to improve reliability and security by design in wireless environments \cite{draft-irtf-icnrg-challenges}, \cite{aso-drmic-13}.
For IoT scenarios, Li et al. \cite{li2014comparative} analyzed that ICN solutions which base forwarding on a global resolution service achieve comparable performance with ICN schemes based on reverse path forwarding (RPF), such as NDN.
Baccelli et al. \cite{baccelli+:2014} showed that ICN can be implemented even on very constrained devices, and that ICN leads to performance gains compared to the currently standardized IoT protocol suite.
To the best of our knowledge, however, there is no work on improving network conditions for the IoT by adapting the \emph{MAC} layer based on principles of RPF-based ICN~solutions.

Amadeo et al. \cite{acm-msdri-14} propose an NDN forwarding engine which allows for reliable multi-source data retrieval in IoT scenarios.
They achieve collision avoidance on the \emph{network} layer as consumers compute a random contention window for transmission.
In this paper, we concentrate on reservation-based approaches on the \emph{link} layer for the sake of robustness and efficiency.
Furthermore, it is worth noting that our approach operates below the network layer, which leads to the following benefits.
First, it abstracts from specific NDN implementations and thus broadens deployment.
Second, it directly controls the duty cycling of the network controller.
This is crucial with respect to energy saving because it enables to switch wireless cells off aligned with data transmission requirements.
It also eliminates radio interference.
TSCH multiplexes in time and frequency.
Having control over the frequency per node is particularly important for IoT scenarios, where no infra\-structure-based controlling of the wireless spectrum can be assumed.



Based on the observation that subsequent data chunks may vary significantly, Arianfar et al. \cite{aso-drmic-13} propose the assignment of an explicit lifetime to ICN packets to improve resource management at nodes as well as within the ICN network.
The lifetime is derived from application requirements. Such information could be used to specify scheduled TDMA more precisely.
However, in this paper we focus on a very basic adaption which does not require additional meta data.
Furthermore, we follow the current implementations of CCN/NDN and thus consider the lifetime meta data as an optional optimization in future~work.

%% file: text/related.tex
\section{Technical Background on IoT}\label{sec:related}

\textbf{Fundamentals of the IoT}
\quad
Over the last decade, the simultaneous availability of (i) low-power radio and MAC protocols such a IEEE~802.15.4, and (ii) small-foot-print, cheap hardware has successively given birth to wireless sensor networks, and the IoT. 
Originally, the IETF has standardized a suite of specifications that adapt IPv6 to memory-constrained nodes and wireless communications characteristics typically encountered in the IoT based on IEEE~802.15.4.
This suite of standard specifications includes dedicated protocols, for example, for routing (e.g., RPL) and HTTP-like communication (e.g., CoAP) \cite{palattella+:2013,sheng2013survey}.
A recent amendment of IEEE~802.15.4 (based on CSMA) is the TDMA-based IEEE~802.15.4e~\cite{802-15-4e-2012}, which has been the subject of intense interest for the IoT community.
To adapt IoT tailored protocols to IEEE~802.15.4e, a new IETF working group~\cite{ietf-6tsch} has been created, called 6TiSCH.
In order to better grasp the technical reasons for the appeal of IEEE~802.15.4e in the IoT, the we recall some basics on wireless MAC protocols that are useful at this stage.

\vspace{2pt}\noindent
\textbf{Contention- vs. Reservation-based Wireless MAC}
MAC protocols can be categorized into (i) contention-based and (ii) reservation-based approaches.
The advantages and drawbacks of each categories have been discussed extensively in the literature (e.g., in \cite{kulkarni2004tdma, cionca2008tdma}). 
Now, we summarize the key points, relevant in the context of this paper.
The advantages of \emph{contention-based protocols} (e.g., CSMA) are their simplicity, and little to no prerequisites.
In particular they do not rely on clock synchronization between nodes or on the availability of network topology information.
Therefore, such protocols are good candidates to cater for dynamic network membership, e.g., in networks of mobile nodes.
Contention-based protocols usually provide satisfactory performance in scenarios with low utilization of the shared medium and sparsely distributed nodes.
One of the main drawbacks of contention-based protocols is their high energy consumption due to idle listening, albeit techniques such as preamble-sampling and duty-cycling can help to mitigate this.
Another drawback is the increasing probability for collisions in case of higher traffic load.
Furthermore, contention-based approaches are likely to suffer from external interference on the communication channel since their unsynchronized nature prevents them from applying channel hopping techniques.
The most prominent example of CSMA-based mechanism for wireless is IEEE 802.11.

In contrast, \emph{reservation-based approaches} (e.g., TDMA) require some knowledge about the network topology and neighbor node's configuration, in order to build a transmission schedule.
This schedule can have various optimization goals, e.g., fairness, latency minimization, or throughput maximization.
TDMA divides time into \emph{timeslots} that may be grouped into \emph{slotframes} or \emph{superframes}.\footnote{For the remainder of this paper we will use the term \emph{slotframe} for a group of timeslots that is repeated over time as defined in the IEEE~802.15.4 specification.}
Nodes know about the schedules of their neighbors, and thus only need to wake up in timeslots that are either reserved for themselves or their neighbors.
This allows to eliminate most of the need for idle listening.
Moreover, the schedule can be designed so that collisions are completely avoided, thus drastically increasing determinism and reliability of the network.
Examples of scheduling techniques for TDMA include the Neighborhood-aware Contention Resolution (NCR) algorithm~\cite{bao2001new}, TRAMA~\cite{rajendran+:2006}, a traffic-adaptive medium access protocol targeting energy-efficient collision-free channel access in sensor networks, and STORM~\cite{garciamenchaca:2012}, a cross-layer framework for disseminating real-time and elastic traffic in multi-hop wireless networks.
The drawbackt this is that nodes have always to listen to the channel when they do not \emph{own} the slot or do not make a reservation.
\oleg{Is this last sentence also true for NCR and TRAMA or only for STORM?}

\vspace{2pt}\noindent
\textbf{Time Slotted Channel Hopping (TSCH)}
\quad
Wireless transmissions in a TDMA-based network may still suffer from interferences. 
These are either (i) external interferences, e.g., caused by another wireless network which is co-located, or (ii) internal interferences, e.g., caused by multi-path fading \cite{watteyne+:2009}.
In order to mitigate the effect of interferences, channel hopping techniques can be used to transmit on multiple channels in a synchronized manner.
Previous work has indicated that TDMA combined with channel hopping (i.e., TSCH) can significantly increase connectivity, efficiency, and stability of a network~\cite{watteyne+:2009,doherty+:2009,duquennoy15orchestra}, achieving up to 99.99\% end-to-end~reliability.
\oleg{Also cite Contiki paper if already published.}

Every node in a TSCH network maintains a schedule (which repeats in every slotframe).
This schedule can be represented as a matrix (timeslots as columns and channel offsets as rows) where cells can be reserved for receiving, sending, or broadcasting (shared cells).
Reserving cells in a TSCH schedule can be done either by using a \emph{node scheduling} algorithm or a \emph{link scheduling} algorithm~\cite{dezfouli+:2015}.
A node scheduling algorithm ensures that each node's transmission timeslot does not conflict with any transmission timeslot of its 1-hop or 2-hop neighbors.
This guarantees that a node's transmission can be received by each of its 1-hop neighbors.
Hence, this is a suitable approach for broadcast transmission.
A link scheduling algorithm guarantees that each transmission of any node to a specific neighbor is receivable by the intended receiver.
This receiver-oriented scheduling is suitable for unicast transmission.
Both types of scheduling techniques serve different purposes and accordingly, a TSCH schedule may include both \emph{broadcast cells} and \emph{unicast cells}.

The cells for a transmission schedule in TSCH are either subject to \emph{static reservations} or to \emph{dynamic reservation}. 
Static reservation allocates the cells once in the beginning and keeps this schedule until the node leaves the network.
In contrast, dynamic reservation allows a node to reserve cells only on demand, e.g. in response to the node's current traffic load.
Cells may be added or removed to the schedule at any time.
However, negotiating and modifying these reservations introduces additional overhead.
Periodic information exchange---either between neighboring nodes or towards a central entity---is necessary to (i) update the information on each node's current neighborhood and schedule and (ii) either a negotiation protocol between neighboring nodes~\cite{zhu2001five} or traffic for requesting and assigning the schedule by a central entity such as PCE~\cite{RFC-4655}, or TASA~\cite{palattella2012traffic}. 
In this context, DICSA provides a distributed and concurrent link scheduling algorithm that requires no specific assumption regarding the underlying network~\cite{dezfouli+:2015}. DeTAS \cite{accettura2013decentralized} provides another distributed link scheduling algorithm specifically targeting 6TiSCH~\cite{ietf-6tsch}. 
Tinka et al. proposed a simple scheduling mechanism for the TSCH MAC protocol that aims for full connectivity with a focus on mobile nodes and a dynamically changing neighborhood~\cite{tinka+:2010}.

TSCH is the core mechanism of common wireless communication standards targeting the IoT domain, such as WirelessHART, ISA100.11a and IEEE~802.15.4e, the latter being our focus here in practice.
While the community puts a lot of work into adapting the IPv6 stack to IEEE~802.15.4e, only very little work is around showing how TSCH might benefit from ICN---which is the topic of this paper.

%% file: text/icn-over-tsch.tex
\section{The Idea of ICN over TSCH}\label{sec:icn-tsch}

\subsection{The Potentials for Link-Layer Adaptation}

\textbf{(1) NDN Traffic Patterns} 
\quad
Content distribution in NDN follows a request/response pattern with footprint on each hop. A request is propagated hop-by-hop in an Interest packet and implements a Pending Interest (PI) state in the corresponding tables (PITs) of intermediate nodes.
Such a PIT entry matches at most one data chunk of limited size. Hence, in a fully deterministic, lossless setting, each request is answered by a train of up to $k$ data packets within a time frame bound by the (temporal) diameter of the network.

For scheduling the wireless, we can interpret an Interest as a predictor of  data expected on the reverse path, and conversely can exclude any data arrival in the absence of PI state. We can further exploit the predefined chunk size for fixing the ratio of data per Interest packet in our schedule. Ideally, the arrival of an Interest would trigger the allocation of $k$ slot frames towards the appropriate neighbor at the expected time. 

However, as explained in \S~\ref{sec:related}, the dynamic reservation of cells requires coordination among neighbors and cannot be efficiently implemented chunk-wise. Neighbor selection furthermore assumes unambiguous routing information in place, which is often exceptional in IoT environments. We will show in the following sections how to procure routing and adapt scheduling in an efficient manner.

\vspace{2pt}\noindent
\textbf{(2) NDN Faces}
\quad
NDN introduces the concept of faces as an abstraction of logical network interfaces between neighboring nodes. Faces map to point-to-point links in a typical wired environment. In low power wireless networks, though, nodes with omnidirectional antennas participate in shared links between a group of neighbors. Neighbor-specific faces (e.g., L2--tunnels) without isolation on links cannot freely co-operate, but will interfere with each other. 

The use of a transmission schedule in TSCH allows to establish a cell-to-face mapping, while each cell (except for broadcast) is assigned to allow (unidirectional) transmission between individual nodes, only. Consequently, all scheduled cells within the transmission matrix of a node can be mapped to the corresponding faces.
\oleg{The term "matrix" in this context should be introduced before.}
Each face (except for a broadcast face) will typically consist of at least two cells---one RX (receive) cell and one TX (transmit) cell.

Frequency division multiplexing in TSCH enables data transmission within multiple cells at the same time. Spreading channels among faces will allow to schedule several faces in parallel. A node can thus be enabled to communicate with several neighbors in the same timeslot.

\subsection{Design Aspects and Requirements}

In our following design, we focus on a typical IoT deployment scenario of a multi-hop wireless network that can reach the Internet via  at least one gateway.
While the nodes may be constrained, the gateway is assumed to have sufficient memory resources for holding a full FIB.
\oleg{What is a \textbf{full} FIB and shouldn't it be RIB?}
Furthermore, we assume a fairly static topology with mostly  stationary nodes, since mobility is not in the focus of IEEE~802.15.4e~\cite{palattella2013optimal}.

A  use of ICN on  TSCH in a network requires the following basic coordinative elements of TSCH  in place.

\vspace{2pt}\noindent
\textbf{Time Synchronization}
\quad
Operating the TDMA transmission schedule in TSCH  requires a synchronisation of clocks within a low millisecond range.
The maximum required precision is mostly derived from the \emph{guard time}, which is, for example, set to 1.5~ms in OpenWSN, the de-facto reference implementation of IEEE~802.15.4e.
Common IoT nodes with  cheap oscillators exhibit a clock drift that can exceed  $~30~ppm$, which poses high requirements on a clock synchronisation protocol. However, the required synchrony in time can be achieved either out-of-band (e.g., using a GPS signal), or by state-of-the-art clock synchronisation protocols for low power networks, such as the Gradient Clock Synchronization Protocol (GTSP) \cite{gtsp-09} or an adaptive synchronization towards a root node in tree-based topologies. 

\vspace{2pt}\noindent
\textbf{Frequency Coordination}
\quad
When calculating a schedule, each node needs to be aware of all neighbors that are in transmission range, so that unwanted overlaps in the time-frequency domain can be reliably avoided. For an efficient scheduling of nodes, a space-frequency division would be obstructive, and hence nodes need also knowledge about their two hop neighborhood. This information should be provided from topology building and used by a  reservation protocol that is needed for negotiating the schedule among neighbors \cite{draft-wang-6tisch-6top}. Both protocols can operate below the network layer and without interfering with NDN.

\subsection{Topology and Routing}

\textbf{(1) Initializing a DODAG}
\quad
For successfully scheduling in frequency and time, we first need to create a topology within the network of IoT nodes. We propose to follow the common approach of building a tree-like structure---a destination-oriented directed acyclic graph (DODAG)---as known from RPL \cite{RFC-6550} with the IoT gateway in the role of the root node. 
Parents broadcast their presence (DIO) and children attach (DAO).
These link-local operations can be transfered to the link-layer in a straight-forward manner. To facilitate frequency coordination, it can also be easily extended to inform about 2-hop neighbors.

Given this basic topology, every node can identify up- and downward paths and thus reach the gateway (root). We now need to address the more delicate question about arbitrary ICN routing on names. Here, we need to face the trade-off that Interests in a scheduled environment best float on a precise paths, but intermediate nodes have limited memory and cannot hold large routing tables. 

\vspace{2pt}\noindent
\textbf{(2) Learning Routes to Names}
\quad
In our previous work \cite{baccelli+:2014}, we have designed and analysed two routing mechanisms---Vanilla Interest Flooding (VIF) and Reactive Optimistic Name-based Routing (RONR). While VIF works without a FIB, RONR nodes gradually acquire FIB entries in a reactive fashion. Given the DODAG topology, we will now follow the PANINI approach \cite{swbw-panii-15}---an optimized strategy for  routing Interests on names that makes a hybrid use of both routing primitives.

We select the gateway as the routing core under the previous assumption that it can hold a full routing table. Every node that offers a routable name advertises this name to the gateway. These Name Advertisement Messages (NAMs) travel hop-by-hop towards the root, and every intermediate node is free to update its own routing table. Intermediate nodes are not required to have a full FIB, but rather aim at adapting a few FIB entries to optimize guidance for Interests. Thus, each node autonomously decides about (a) its memory resources dedicated to the FIB, and (b) the forwarding logic it applies within its vicinity. Traffic flows can be continuously used to adapt the FIB to relevant traffic patterns. For example, a node can hold more specific information for frequently requested names, while it may erase entries for rather unknown traffic.     

\vspace{2pt}\noindent
\textbf{(3) A Bimodal FIB}
\quad
The objective of the FIB at intermediate nodes is to optimize traffic flows at minimal storage cost. For this, we propose to extend the FIB structure to hold two modes---{\tt include} and {\tt exclude}. In {\tt include} mode, all Interests that match a FIB prefix will be forwarded on the associate Face, while all Interests that match a FIB {\tt exclude}-prefix will be blocked on that Face. The initial state of an empty FIB reads {\tt include~*} which leads to a transparent forwarding (flooding) of all incoming Interests. A node that has seen no routable names from NAMs in a subtree of his may as well switch to {\tt exclude~*}. Based on this bimodal mechanism, typical optimizations could be as follows. An intermediate node sees much traffic of names with a prefix {\tt /light/*} from many of its children, but some subtree(s) does not provide  {\tt /light/}-data. Assigning a single {\tt exclude /light/*} to the corresponding Face(s) may result in an efficient trade-off between FIB memory and unwanted Interest traffic. A particularly effective optimisation can take place, if a node knows about {\tt /light/*} in downward direction. It can place  {\tt exclude /light/*} at the upstream keeping all corresponding traffic local. 

It is noteworthy that in the machine-to-machine oriented setting of the IoT it is easier to arrange names and topology in an aggregatable fashion, so that short prefixes may be effective  for large collections of IoT data~sources.

\vspace{2pt}\noindent
\textbf{(4) Routing to Names}
\quad
After initial NAMs have arrived at the gateway and in the absence of any distributed routing knowledge, all nodes can reach all names by transmitting the Interest upwards. If an Interest cannot be satisfied on path, it will travel upwards to the root node, where it is flooded down  its proper subtree. Even though suboptimal, this default routing is surprisingly lean, as we will discuss in Section \ref{sec:evaluation}. Note that every node throughout the network can always tell whether an interest travels upwards, or downwards and thus can restrict flooding.

In the presence of meaningful, distributed FIBs, both routing phases benefit from optimization. Each hop on the upward path can redirect an Interest downwards to a local subtree, if a matching FIB entry exists. In the downward flooding phase, every request-related FIB entry  narrows the dissemination of an Interest, and in the ideal case leads to a unique shortest path to the named data provider. It is worth recalling that nodes can adapt routing precision to traffic patterns so that frequently requested names or prefixes become more present in relevant FIBs.  

%% file: text/scheduling.tex
\section{Scheduling}\label{sec:scheduling}

\newcommand{\SInt}{$SSF_{I}$\xspace}
\newcommand{\SCont}{$SSF_{C}$\xspace}
\newcommand{\SDyn}{$SSF_{Dyn}$\xspace}

We now describe the design of a schedule for TSCH that is compliant to the ICN traffic pattern and adaptive to data demands.
This shall flexibly optimize network performance and minimize energy consumption, but must not increase  complexity for node coordination (see \S~\ref{sec:icn-tsch}).

The general idea is a schedule that is partly static and pre-reserved, and partly dynamic and adaptive to the current traffic pattern. For this, we divide the slotframe into three parts, henceforth called  subslotframes (\emph{SSF}s). The first  \emph{SSF} is dedicated to statically scheduled Interest propagation and named  \SInt{}. Second, \SCont{} is for sending back content chunks on a semi-dynamic schedule. The schedule of the third \emph{SSF} is fully dynamic. This  \SDyn{} is activated to serve increased traffic loads on dedicated~links.

For the following description of the scheduling procedure, we define $G = (V, E)$ as an undirected graph with a set of vertices $V$ representing the set of nodes and a set of edges $E$ representing the links between two nodes present in the routing graph.
If two nodes $a$ and $b$ share an edge $(a, b) \in E$, they are called \emph{1-hop neighbors}.

\vspace{2pt}\noindent
\textbf{$\mathbf{SSF_{I}}$ -- Static Interest Schedule}
\quad
The cells in this first subslotframe are reserved at network bootstrapping after the topology is created (or reconfigured).
For reconfiguration purposes, the reservation of the first cell ($c(1,1)$) is fixed to a general broadcast (of entire wireless range) and used to alert all nodes within wireless reach. \oleg{Introduce this notation in section 3.}
Nodes that do not need to send any reconfiguration data, are required to switch to receiving mode for slot $1$ at channel offset $1$.
Each node reserves a predefined number of TX cells to each of its 1-hop neighbors, and a matching RX cell (same slot number, same channel offset) for each  TX cell a 1-hop neighbor has allocated towards it. In this way, basic capacities for exchanging Interests among neighbors are defined.
    The amount of reserved cells per neighbor can be chosen according to a priori knowledge of communication patterns---upstream (or default) routes may receive higher capacities, for example.

Additionally, a node should reserve cells for broadcasting to cope with incomplete routing information. Broadcast capacities may be aligned with predictable traffic patterns and available FIB memory. Interest broadcasts are limited to  1-hop neighbors and different from the general broadcast in cell $c(1,1)$.

\vspace{2pt}\noindent
\textbf{$\mathbf{SSF_{C}}$ -- Semi-dynamic Content Schedule}
\quad
Each Interest is potentially answered by a content chunk.
Taking this information into account and assuming a maximal chunk size of $k$ packets, the content schedule in the second \emph{SSF} shall be built as follows.
For each RX cell in \SInt{}, a node reserves $k$ TX cells, and for each TX cell in \SInt{}, a node reserves $k$ RX cells.
As such, the cell assignment  does {\em not} require any negotiations between nodes, but is  a direct consequence of the  \SInt{}, and static. 

However, the  nature of NDN traffic allows for an adaptive operation of the \SCont.
Initially,  all reserved cells are deactivated, which means that the transceiver will not be switched on and the CPU may remain in energy saving mode. 
Node $b$ activates $k$ RX cells for a neighboring node $a$, after an Interest has been sent to $a$ in \SInt{}.
These cells will get deactivated again, either after a content chunk was received from $a$, or when the PIT entry times out and is removed.
By deactivating cells, energy can be saved from reducing idle listening and increasing the time the CPU can spend in sleep mode.

In the case of Interest broadcasting, these savings cannot apply. To limit broadcast reception periods, we assign shared cells to \SCont. A TSCH shared cell operates CSMA/CA for increased flexibility at the price of reduced reliability.   

\vspace{2pt}\noindent
\textbf{$\mathbf{SSF_{Dyn}}$ -- Dynamic On-Demand Schedule}
\quad
Cells in the third part of the slotframe stay unreserved at bootstrapping, and are only activated if traffic demands exceed the initially foreseen capacities. On a per link base, a balanced set of Interest and content cells are (de)allocated dynamically between two nodes and adapt the wireless spectrum to current utilization patterns. In detail, each node monitors the utilization of the (directional) links to each of its neighbors.
Link utilization $U$ is measured as the ratio between \emph{used} cells $c_{u}$ and \emph{scheduled} cells $c_{s}$: $U = c_{u}/c_{s}$ .
\oleg{Was the definition of used and scheduled cells removed on purpose?}

If the recent link utilization $U_{cur}$ from node $a$ to node $b$ over a pre-defined time period $T$ exceeds a predefined threshold $U_{Th}$, $a$ and $b$ reserve a preconfigured set of additional slots for sending/receiving Interests and content in $SSF_{Dyn}$. Thresholds and allocated slot sizes are parameters of the network that can be adjusted to meet deployment-specific criteria (see example below).
Deallocation is performed after the $U_{cur}$ falls below a certain threshold $U_{Tl}$ in $T$.
In this way, radio resources can be dynamically adapted to actual (bursty) traffic demands that may vary between node pairs, while low (regular) communication requirements allow for extended sleeping cycles in radio interfaces and thus enhance energy efficiency.

The dynamic adaptation of the schedule requires coordination between 1-hop and 2-hop neighbors. The information  about a node schedule and the schedule of its 1-hop neighbors can be piggy-backed in ICN (Interest) traffic in a memory-efficient representation (such as bit fields).
In this manner, a node will gain knowledge about the schedules of all nodes within its 1-hop and 2-hop neighborhood.
This information serves as basis for reserving additional cells in \SDyn{} by a link scheduling protocol like \emph{LAMA}.

\vspace{2pt}\noindent
\textbf{Example}
\quad
Assuming a typical building automation scenario nodes may request (period) configuration and software updates---e.g., provided by gateway acting as the root node ($1$) in the routing tree.
Taking this knowledge into account, nodes will make more reservations in \SInt for upstream packets.
Let a slotframe consist of $101$ slots (as proposed by the IETF 6TiSCH WG) and 16 channel offsets (according to the 16 channels available in IEEE~802.15.4).
For simplicity we assume furthermore that $k = 1$.
A sensible partitioning could be to assign 20~slots to \SInt and \SCont respectively.
Depending on the network's density a node may reserve 1 (high density) to 9 (very low density) cells per neighbor in each of the first two SSFs.
The remaining 60 slots---remember that the first slot is reserved for broadcasting---are assigned to \SDyn and thus unreserved in the beginning.
While the cells reserved in \SInt and \SCont may suffice \oleg{check word} the general requirements for fetching and delivering configuration information, it may happen from time to time that more data has to be delivered to the downstream nodes, e.g. in case of a firmware update.
In this case, nodes will detect a high utilization of the cells in \SInt and \SCont and according make reservations for these links in \SDyn. 
Hence, up to 30 additional cells may be reserved for Interests and content chunks respectively.
After the firmware update is fully delivered to the affected \oleg{check word} nodes, reservations in \SDyn can be deallocated again.

It can be seen that the sizes of ideally \SInt{} and \SCont{} should be kept considerably small and only ensure basic connectivity, in order to assign more cells to \SDyn{}.

%% file: text/evaluation.tex
\section{Evaluation}\label{sec:evaluation}

\subsection{Routing Analysis}

The proposed routing mechanism leads to forwarding on shortest paths, provided all nodes hold full FIBs. Our evaluation shall concentrate on the analysis of worst cases, when FIBs of intermediaries are empty and flooding is required. More precisely, we consider the cases where flooding starts at some intermediate node and continues throughout the subtree (i.e., no subsequent forwarder has additional FIB~knowledge).   

The number of  nodes in a `vanilla--flooded' subtree strongly relies on the topology, and we seek the distributions of node counts over all possible subtrees involved. This cannot be evaluated on our testbed, where topological variations are confined to the physical setting. Hence we perform this topological study based on rigorous theory and on simulations.

The RPL-like tree building mechanism generates shortest paths trees, which are theoretically well described by Uniform Recursive Trees (URTs) \cite{m-pacns-06}. Similar to the actual signaling process, a URT is generated by randomly adding new nodes to the existing tree. 

We consider a tree (network) of $N$ nodes that are numbered in the order of attachment. Let $D_N(k)$ denote the number of descendants of node $k > 1$, then the distribution reads \cite{pp-lnitr-07}

\begin{equation}
 {\cal P}\left(D_N(k) = j\right) = \frac{(k-1) \cdot (N - k - j + 1)^{\overline{j}}}{(N - j - 1) \cdot (N - 1)^{\overline{j}}},  
\end{equation}
\noindent with $(n)^{\overline{j}}$ the $j$-th rising factorial power of $n$.\\ 
Summing over all nodes $k$ with equal probability $1/N$ yields the distribution $D_N$ of nodes in a subtree rooted at an arbitrarily chosen node.

\begin{figure} 
\subfigure[Analytical]{\includegraphics[width=0.509\columnwidth,natwidth=100,natheight=100]{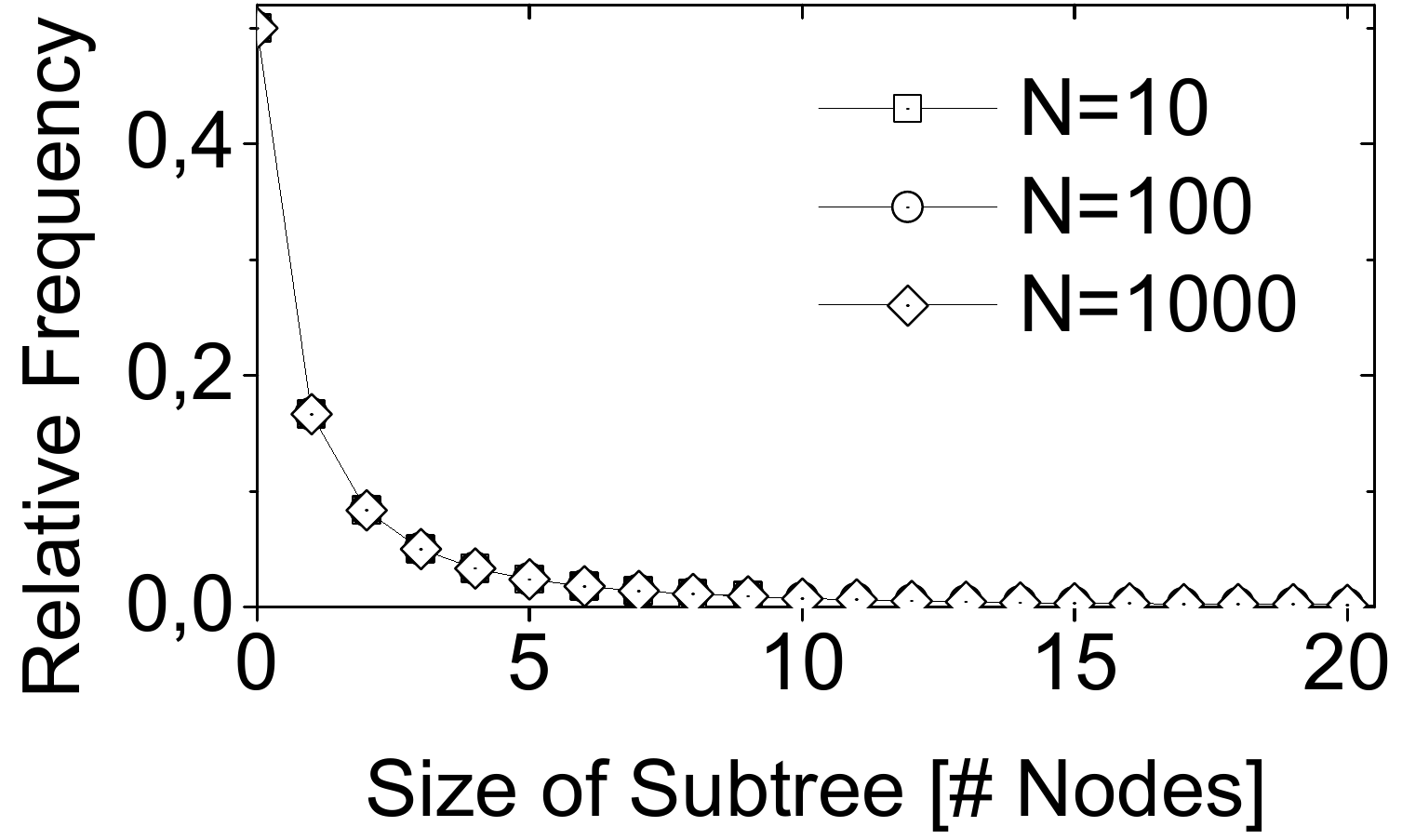}\label{fig:tree-result-ana}}
\subfigure[Simulated]{\includegraphics[width=0.472\columnwidth,natwidth=100,natheight=100]{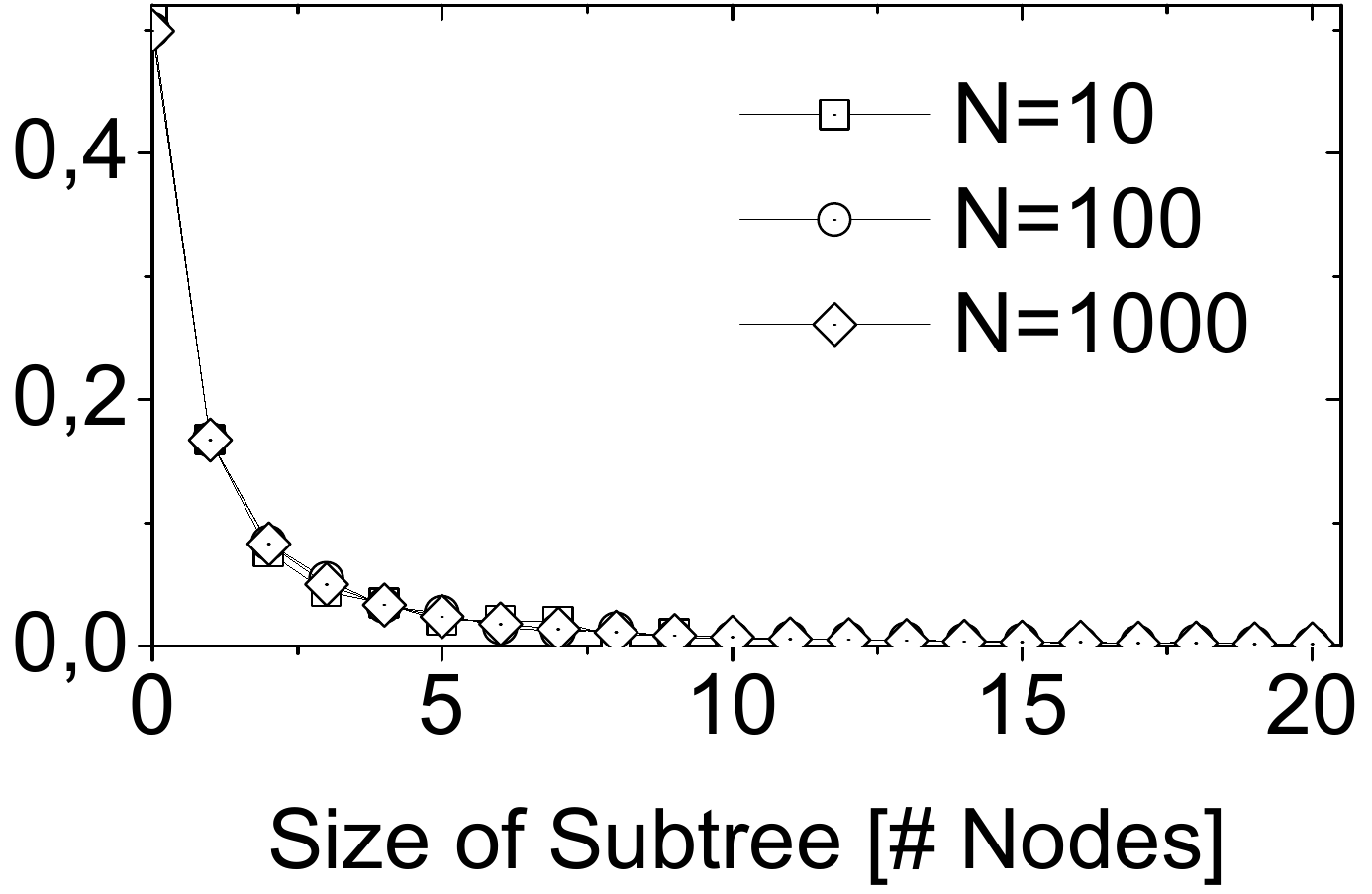}\label{fig:tree-result-num}}
  \caption{Distributions of branch sizes in a \mbox{routing} tree of $N$ nodes}
     \vspace{-1.0em}
  \label{fig:tree-result-num-ana}
\end{figure}
   
Figure \ref{fig:tree-result-ana} visualizes these analytical distributions for different numbers of nodes. Strikingly, the branch sizes are largely independent of the overall network size, which is due to the recursive nature of the URTs.  Node numbers from these exponentially decaying distributions are rather small: more than 7 nodes appear with probability  $0.01$. This is due to a uniformly wide fan-out---trees are rather wide than tall.

Figure  \ref{fig:tree-result-num} displays the same distributions for simulated networks. In these simulations, node topologies were created by connecting new nodes to random parents. Averages have been taken over  $10,000$ iterations.
Both results are in close agreement and support the correctness of the model.

From this brief analysis, we can conclude that non-systematic, random defects in the distributed FIB tables of intermediate nodes have a rather limited impact. In particular, the size of flooded regions does not grow with the total network size. 

\subsection{Scheduling Experiments}
\begin{figure}[t]
    \centering
    \includegraphics[width=0.75\linewidth,natwidth=100,natheight=100]{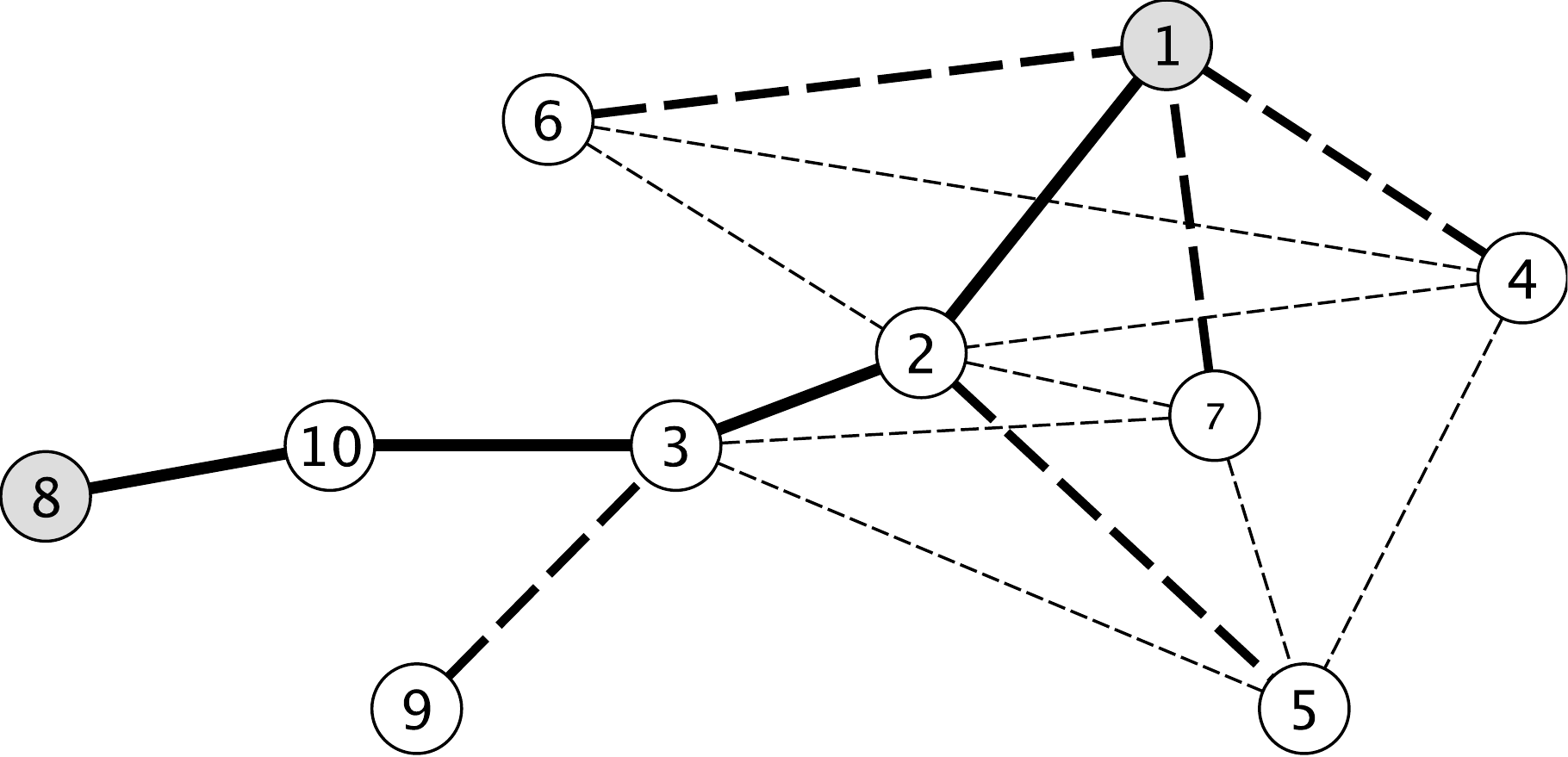}
    \caption{
        Testbed topology in the experiments.
         A connection between two nodes indicates that this link has been scheduled in TSCH.
         Thick lines are part of the formed DODAG.
     }
     \vspace{-1.8em}
     \label{fig:actual-topo}
\end{figure}

\textbf{Experiment Setup}
\quad
In order to evaluate our scheduling solution (see \autoref{sec:icn-tsch} and \autoref{sec:scheduling}), we conducted experiments in the FIT IoT-LAB~\cite{fambon2014fit}.
We compare the approach with an implementation that runs ICN directly on the link layer, using CSMA as a MAC protocol as discussed in our previous work~\cite{baccelli+:2014}.
The hardware platform consists of typical IoT \emph{Class~2} devices~\cite{RFC-7228}, M3 nodes featuring an ARM Cortex-M3 microcontroller with 64~kB RAM and 512~kB Flash, which are equipped with a IEEE~802.15.4 compliant radio.
The software is based on the de-facto standard implementation of IEEE~802.15.4e, OpenWSN~\cite{OpenWSN}, and the open-source IoT operating system RIOT~\cite{RIOT}.

Ten nodes are chosen, forming a multi-hop topology shown in \autoref{fig:actual-topo}.
Node~8 acts as the consumer and node~1 as the content provider as well as the root node in the routing tree.
The requested content consists of 100~chunks.
We assume side traffic from nodes connecting to a (sub)networks.
Therefore, nodes 4, 6, and 7 are also generating traffic with a similar rate as the content consumer (node~8).

\vspace{2pt}\noindent
\textbf{MAC Configurations}
\quad
The static schedule and the routing tree were computed beforehand.
The static schedule for $SSF_{I}$ and $SSF_{C}$ ensures basic connectivity and reserves one cell per link and direction.
We use a length of 15~ms for the slot length and 101~slots per slotframe.
The remaining cells in the slotframe are left initially unscheduled and can be reserved in $SSF_{Dyn}$.
The schedule is constructed in a way that packets can travel from any node along the tree to the root node and from the root node to any node within one slotframe in $SSF_{I}$ and $SSF_{C}$ respectively.
Time synchronization between the nodes is done based on periodic broadcasting of enhanced beacons in shared cells.

For the first series of experiments all scheduled cells in $SSF_{I}$ and $SSF_{C}$ were active and node~8 was sending out Interests with a constant rate of one Interest per slotframe.
We refer to this configuration as \emph{SINR} (Static Information-centric Networking Reservation).

In the second series of experiments the scheduled cells in $SSF_{C}$ were kept initially inactive.
Again, node~8 was sending out Interests with a constant rate of one Interest per slotframe.
As soon as a node~A on the path from 8 to 1 receives an Interest, it activates its RX~cell(s) in $SSF_{C}$ on the link to the next hop B on the path.
Once, A receives the corresponding content chunk from B it deactivates the cell again.
We refer to this configuration as \emph{DINR} (Dynamic Information-centric Networking Reservation).

The next series had the same configuration as for \emph{DINR}, but made also use of the dynamic part of the schedule $SSF_{Dyn}$.
If a node~A receives a certain amount of Interests from one of its neighbors, they implicitly activate cells in $SSF_{Dyn}$ in both directions to increase the bandwidth on this link.
If the cells for this link are less frequently used, the additional cells are deactivated again.
In this configuration we increased the rate in which node~8 generates to 15~Interests per slotframe.
We refer to this configuration as \emph{ADINR} (Adaptive Dynamic Information-centric Networking Reservation).

We compared the results for \emph{SINR}, \emph{DINR}, and \emph{ADINR} with different configurations of ICN on top of a CSMA MAC protocol.
In all configurations, a node initiates up to three link layer retransmissions if no ACK is received.
Interests are retransmitted after a timeout of 1~second by node~8 if no content chunk has been received.
In the first configuration, simply referred to as \emph{CSMA}, node~8 retransmits Interests until it has received the whole content.
The second configuration, referred to as \emph{CSMA-3}, limits the number of Interest retransmissions to three tries.
The last configuration, referred to as \emph{CSMA-3ST}, is similar to \emph{CSMA-3}, but with increased traffic from nodes 4, 6, and 7.

Each serie of experiments is sampled with the same parameter settings until it is converged.

\vspace{2pt}\noindent
\textbf{Results}
\quad
We considered four different metrics: (i) time to completion, (ii) jitter, (iii) end-to-end packet delivery ratio (PDR), and (iv) energy consumption.

\begin{figure}[t]
    \centering
    \subfigure[Time to Completion]{\includegraphics[width=0.49\linewidth,natwidth=100,natheight=100]{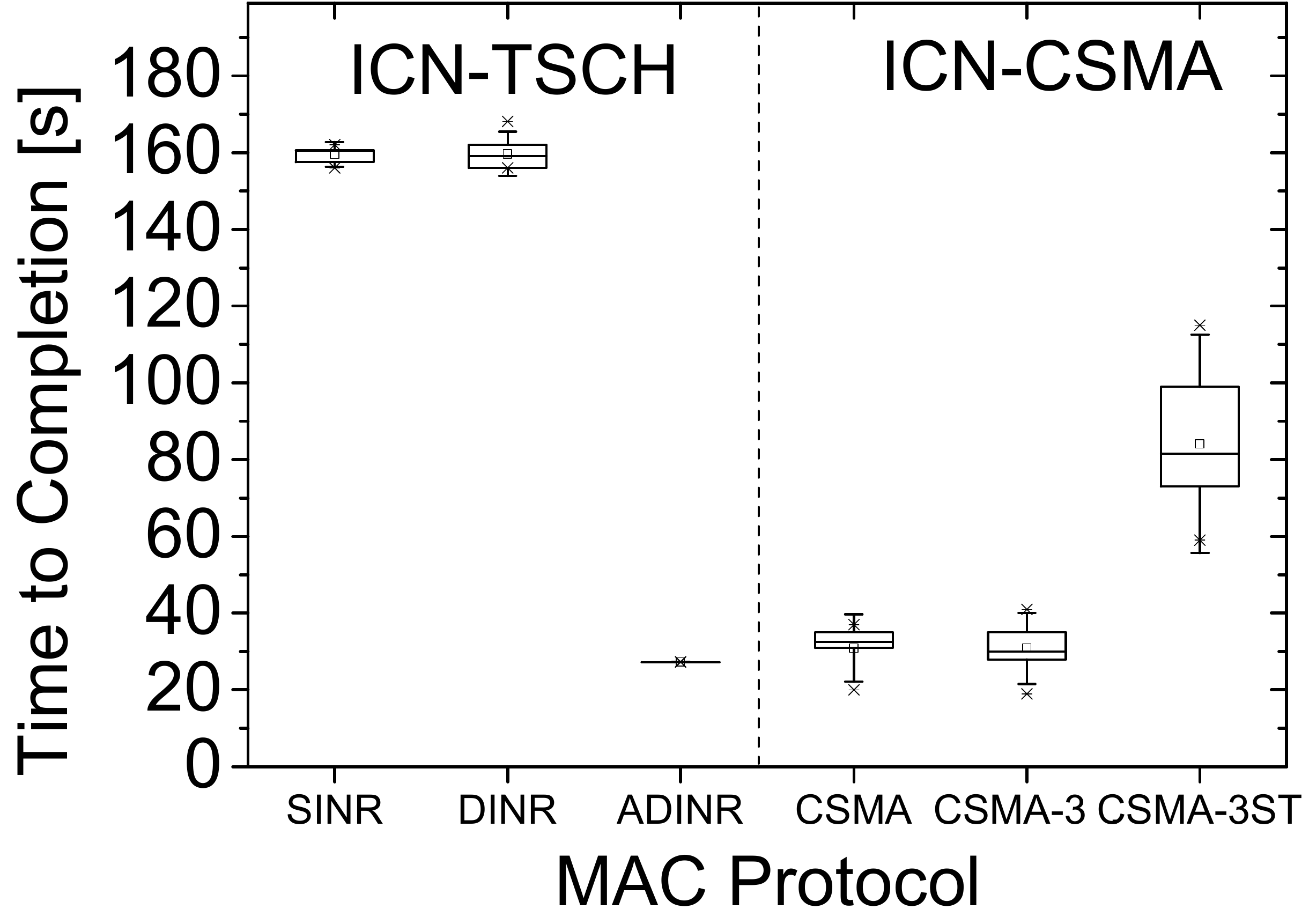}\label{fig:exp-durations}}
    \subfigure[Packet Delivery Ratio]{\includegraphics[width=0.49\linewidth,natwidth=100,natheight=100]{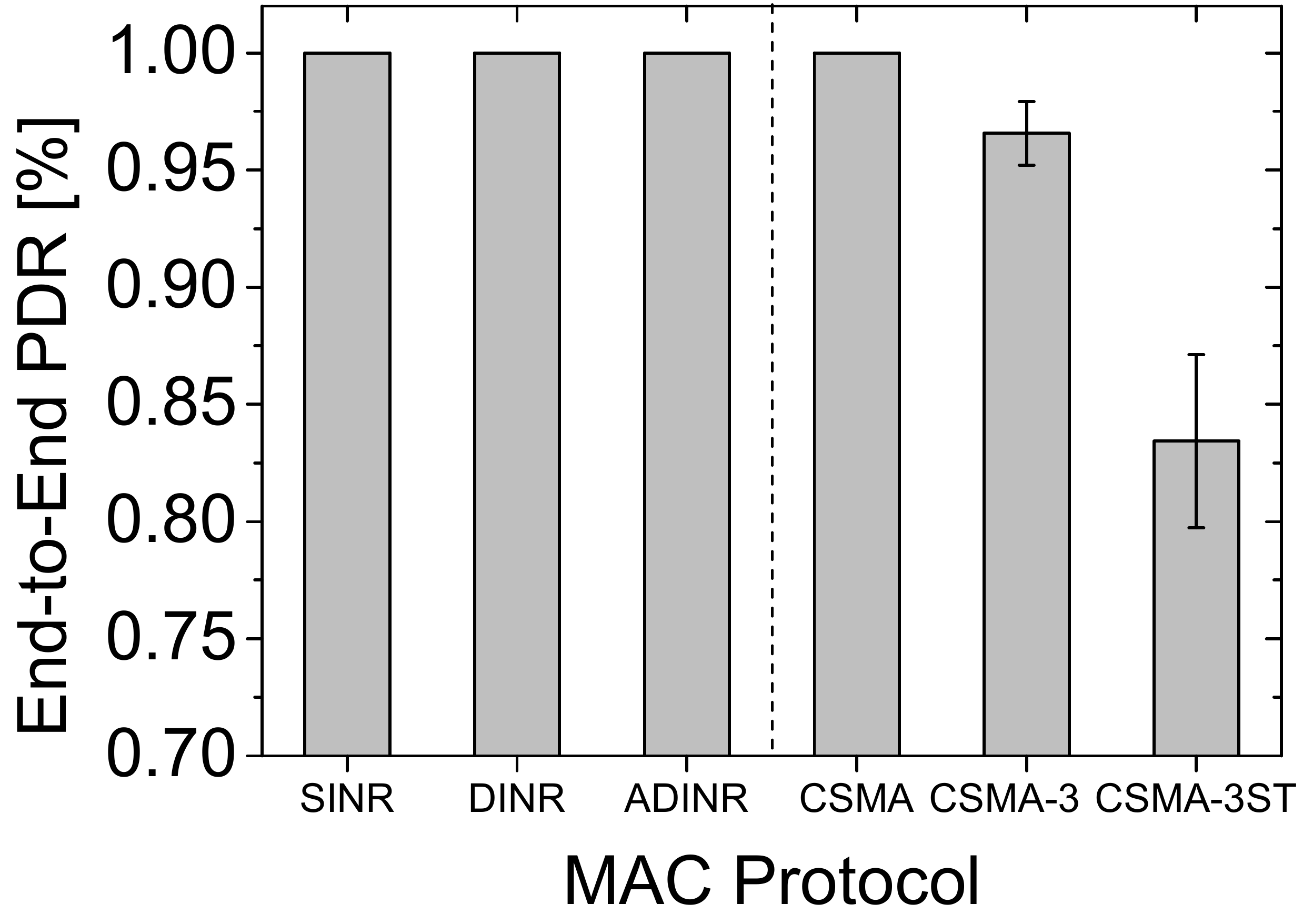}\label{fig:exp-pdr}}
    \caption{
        Comparison of time to completion and PDR in different configurations for TSCH and CSMA.
    }
     \vspace{-1em}
\end{figure}

Since only one Interest and one content chunk can be transmitted per slotframe with \emph{SINR} and \emph{DINR}, the minimum time to completion for fetching 100 content chunks is $\Delta = 100 * T_{SF}$ with $T_{SF}$ being the duration of the slotframe.
As we can see in \autoref{fig:exp-durations}, the measured time is only slightly above this minimum.
Initially, \emph{ADINR} generates more Interests per slotframe than it can send out, but gradually, nodes along the path activate more cells in $SSF_{Dyn}$.
$SSF_{Dyn}$ contains 70~cells which implies that up to 8 additional links per hop (4 hops, bidirectional) can be scheduled.
This leads to a tremendous improvement of the time to completion in comparison to \emph{SINR} and \emph{DINR}.
Concerning jitter, our measurements in \autoref{fig:exp-durations} show a very small standard deviation for SINR, DINR, and ADINR, as expected for a reservation-based~MAC.

In comparison, time to completion with CSMA is much less predictable and depends heavily on the number of collisions and retransmissions (per link and end-to-end).
We observe significantly bigger standard deviation and increasing average if side traffic increases

As expected with a collision-free TSCH schedule, we observed almost no link layer retransmissions with \emph{SINR}, \emph{DINR} and \emph{ADINR} (less than 5 retransmissions overall).
Consequently these mechanisms achieved an end-to-end PDR of 100\% for all three TSCH configurations, as shown in \autoref{fig:exp-pdr}.
With \emph{CSMA}, Interests are retransmitted as many times as required, and thus end-to-end PDR reaches 100\% too, but at the cost of many retransmissions and duplicates.
On average, we counted more than 130~end-to-end retransmissions and 25~duplicate chunks that arrived at the consumer.
Limiting the number of end-to-end retransmissions to three (in \emph{CSMA-3}, see \autoref{fig:exp-pdr}) decreases the PDR to about 97\%, with similar numbers for retransmissions and duplicates.
If side traffic increases (in \emph{CSMA-3ST}) the PDR drops further down, with significantly more retransmissions and duplicates.

\begin{figure}[t]
    \centering
    \subfigure[TSCH]{\includegraphics[width=0.95\linewidth,natwidth=100,natheight=100]{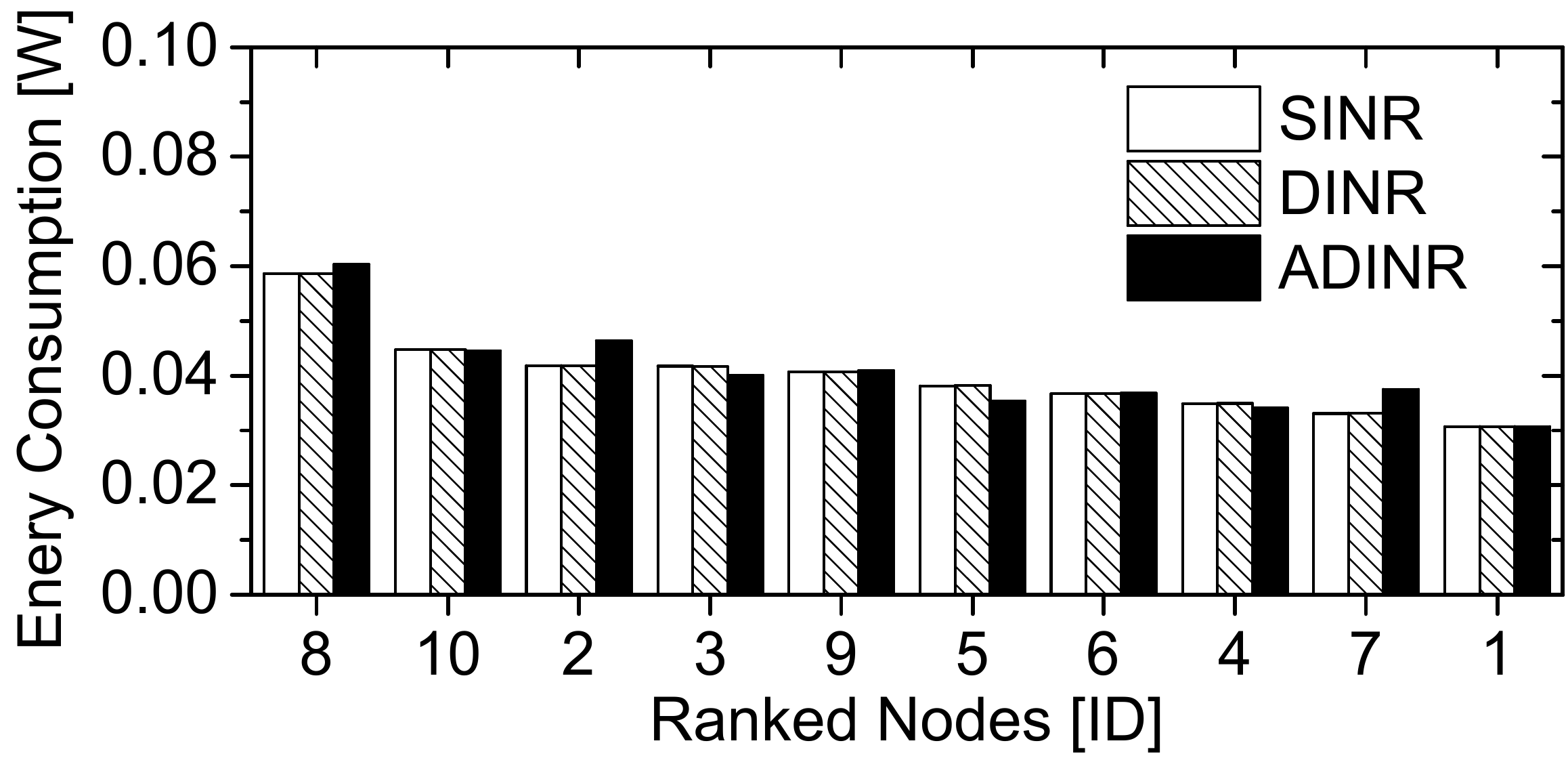}\label{fig:exp-energy-tsch}}
    \subfigure[CSMA]{\includegraphics[width=0.95\linewidth,natwidth=100,natheight=100]{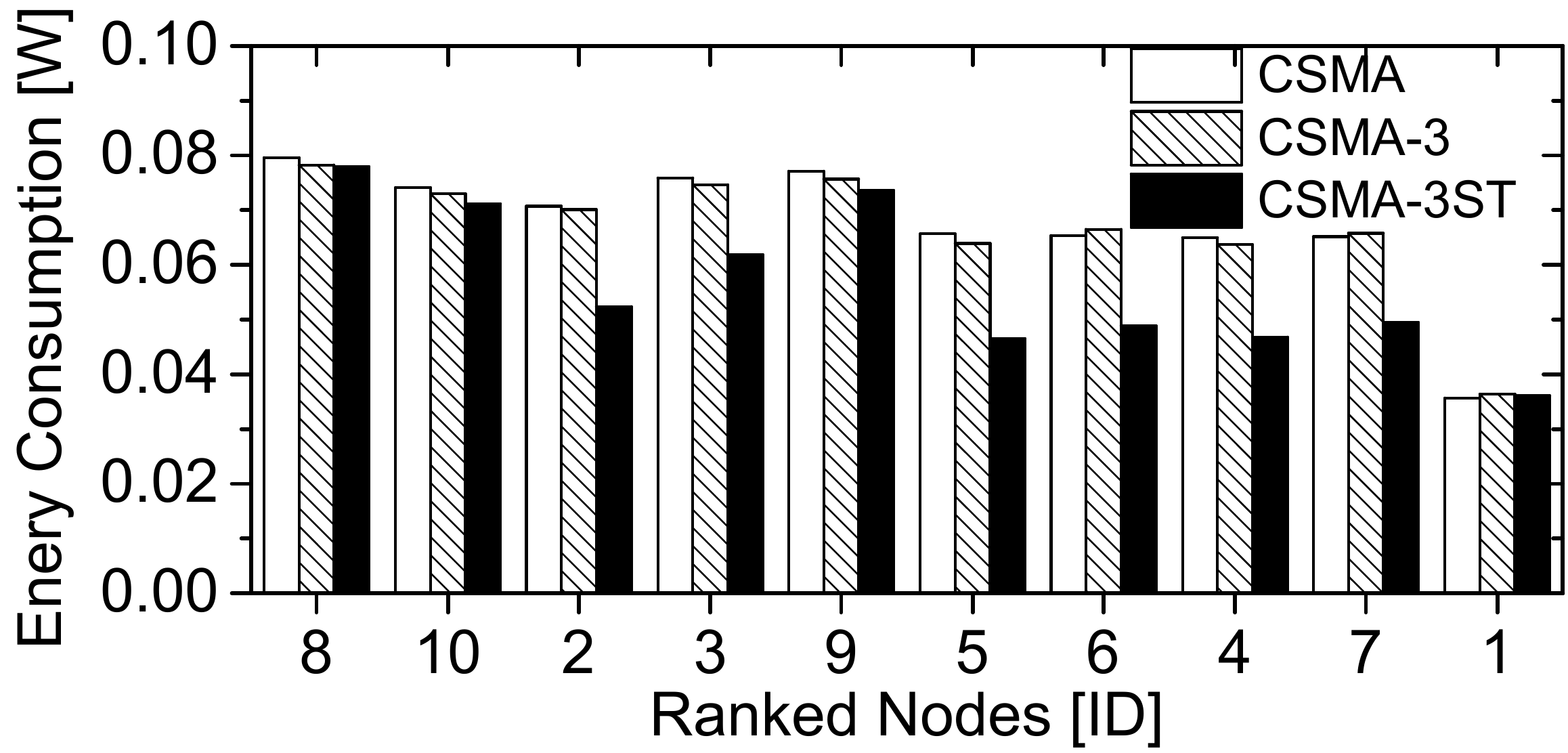}\label{fig:exp-energy-csma}}
    \caption{
        Energy consumption for the different configurations of TSCH and CSMA.
    }
     \vspace{-1em}
    \label{fig:exp-energy}
\end{figure}

The energy measurements were performed using the control nodes provided by FIT IoT-LAB (power consumption measurement through resistor shunts and an INA226 current/power monitor component).
We configured the INA226 with a conversion time of 8244~ms and the averaging mode to 1024 which gives maximum accuracy according to the hardware datasheet.
We computed the average over all samples in all experiment runs, per node, as shown in \autoref{fig:exp-energy}.
With TSCH, transceivers switch to sleep mode for all unscheduled slots. 
Thus, we can see that all nodes consume consistently less power with \emph{SINR}, \emph{DINR} and \emph{ADINR} than with \emph{CSMA}.
Furthermore, we observe that the increased energy consumption in \emph{ADINR} due to a higher duty cycle is leveled out by the fact that the nodes can more quickly return to sleep mode again.